\documentclass[10pt,conference]{IEEEtran}
\IEEEoverridecommandlockouts
\ifCLASSINFOpdf
\else
\fi
\usepackage{amsmath,amssymb,amsfonts}
\usepackage{algorithmic}
\usepackage{array}
\usepackage{graphicx}
\usepackage{mathtools}

\usepackage{textcomp}
\usepackage{subfigure}
\usepackage{caption}
\usepackage{enumitem}
\usepackage{xcolor}
\usepackage{graphicx}
\usepackage{tikz}

\usepackage{bbm}
\usepackage{verbatim}
\usepackage{graphicx}
\usepackage{cite}
\usepackage{url}
\usepackage{amsmath}
\usepackage{color}
\usepackage{cite}
\usepackage{epstopdf}
\usepackage{setspace}
\usepackage{calc}
\usepackage{array}
\usepackage{multirow}
\usepackage[printonlyused]{acronym}
\usepackage{cleveref}
\usepackage{siunitx}
\usepackage{mathtools}

\definecolor{myRed}{RGB}{179, 0, 0}
\definecolor{myViolet}{RGB}{119, 0, 60}
\definecolor{myBlue}{RGB}{0, 0, 77}
\definecolor{myPurp}{RGB}{153, 51, 102}
\definecolor{myOrn}{RGB}{255, 117, 26}
\definecolor{myGreen}{RGB}{18, 157, 41}

\DeclareSIUnit{\belmilliwatt}{Bm}
\DeclareSIUnit{\dBm}{\deci\belmilliwatt}
\DeclareSIUnit{\isotropic}{Bi}
\DeclareSIUnit{\dBi}{\deci\isotropic}
\acrodef{mmW}{millimeter-wave}
\acrodef{BS}{base station}
\acrodef{UE}{user equipment}
\acrodef{SOTA}{state of the art}
\acrodef{AoA}{angle of arrival}
\acrodef{AoD}{angle of departure}
\acrodef{AWV}{antenna weight vector}
\acrodef{ADC}{analog-to-digital converter}
\acrodef{BB}{baseband}
\acrodef{RSRP}{reference signal received power}
\acrodef{CSI}{channel state information}
\acrodef{COTS}{commercial-off-the-shelf}
\acrodef{PAA}{phased antenna array}
\acrodef{TTD}{true-time-delay}
\acrodef{LoS}{line-of-sight}
\acrodef{NLoS}{non-line-of-sight}
\acrodef{IA}{initial access}
\acrodef{DFT}{discrete Fourier transform}
\acrodef{UDN}{ultra-dense networks}
\acrodef{RF}{radio frequency}
\acrodef{MPC}{multipath component}
\acrodef{BF}{beamforming}
\acrodef{SNR}{signal-to-noise ratio}
\acrodef{SINR}{signal-to-interference-plus-noise ratio}
\acrodef{OFDM}{orthogonal frequency-division multiplexing}
\acrodef{DSP}{digital signal processing}
\acrodef{LUT}{lookup table}
\acrodef{MIMO}{multiple-input multiple-output}
\acrodef{IC}{integrated circuits}
\acrodef{PS}{phase shifter}
\acrodef{DAC}{digital-to-analog converter}
\acrodef{EVM}{error vector magnitude}
\acrodef{CP}{cyclic prefix}
\acrodef{FPGA}{field programmable gate arrays}

\newcommand{\hermitian}[0]{\text{H}}
\newcommand{\transpose}[0]{\text{T}}

\begin{document}
%
\title{Structured Two-Stage True-Time-Delay Array Codebook Design for Multi-User Data Communication}

\author{\IEEEauthorblockN{Aditya Wadaskar, Ding Zhao, Ibrahim Pehlivan, and Danijela Cabric}\\ \vspace{-1mm}
\IEEEauthorblockA{Department of Electrical and Computer Engineering, University of California, Los Angeles\\
Email: adityaw@ucla.edu, dingzhao99@ucla.edu, ipehlivan@ucla.edu, danijela@ee.ucla.edu}
\vspace{-7mm}
}


\vspace{0mm}

\maketitle


\begin{abstract}
Wideband millimeter-wave and terahertz (THz) systems can facilitate simultaneous data communication with multiple spatially separated users. It is desirable to orthogonalize users across sub-bands by deploying frequency-dependent beams with a sub-band-specific spatial response. True-Time-Delay (TTD) antenna arrays are a promising wideband architecture to implement sub-band-specific dispersion of beams across space using a single radio frequency (RF) chain. This paper proposes a structured design of analog TTD codebooks to generate beams that exhibit quantized sub-band-to-angle mapping. We introduce a structured \textit{Staircase TTD} codebook and analyze the frequency-spatial behaviour of the resulting beam patterns.  We develop the closed-form two-stage design of the proposed codebook to achieve the desired sub-band-specific beams and evaluate their performance in multi-user communication networks.

\end{abstract}


\section{Introduction}
\label{sec:Introduction}
\begin{figure}[t]
        \centering
        \includegraphics[width=\linewidth]{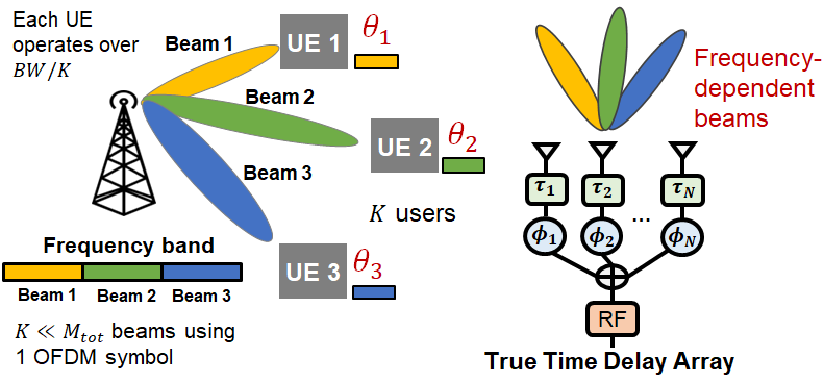}
        \caption{\small Sub-band-specific beamforming for simultaneous multi-user data communication with analog True-Time-Delay arrays.}\vspace{-4mm}
        \label{fig:Introduction}
    \end{figure}

     Millimeter-wave and terahertz (THz) systems offer large bandwidths\cite{Intro_largeBW,7999294,6824752} which, besides enabling high data rates, can facilitate simultaneous data communication with multiple spatially separated users occupying non-overlapping sub-bands. To support such sub-band-specific data communication, base stations need to deploy directional beams with a sub-band-specific spatial response, where all frequency resources within a sub-band form a beam to serve a particular user\cite{Samsung_1,UCSD}, as shown in Fig. \ref{fig:Introduction}.
    While the conventional analog phased arrays can only generate frequency-flat spatial responses, fully digital or hybrid analog-digital arrays that leverage multiple RF chains for enhanced beamforming capabilities incur high costs and power consumption. 

    True-Time-Delay (TTD) arrays are a promising candidate for sub-band beamforming owing to their low-complexity implementation of frequency-dependent beams using a single RF chain. Works in \cite{yan_ttd_2019,9154233,3D_rainbow,vb_ttd_2021} use analog TTD arrays to implement a fully dispersive rainbow beam codebook 
    scanning a continuous range of angles for expedited beam training. Recent works, namely Joint-Phase-Time-Arrays (JPTA)\cite{Samsung_1} and mmFlexible\cite{UCSD}, leverage analog TTD-inspired architectures to generate beams with quantized sub-band-specific dispersion in space. 
    The algorithm proposed in \cite{Samsung_1} iteratively optimizes the per-antenna delays and phase shifts, whereas the algorithm in \cite{UCSD} is based on a closed-form Least-Squares solution. {\let\thefootnote\relax\footnote{{{This work was supported by the NSF under grant 1955672.}}}}
    
    In contrast with \cite{Samsung_1,UCSD}, this paper adopts a structured beam-synthesis methodology rooted in principles of array design and frequency-spatial beam-pattern analysis to design sub-band beams, rather than target-based optimization or pattern-fitting.
    The main contributions of the paper are summarized as follows: 
    We propose a structured delay-phase codebook called \textit{Staircase TTD} codebook in Sec. \ref{sec:system_model}, and study the frequency-spatial characteristics of resulting beams in Sec. \ref{sec:III}. 
    We then develop 
    a closed-form design of the proposed codebook to implement dual-stage frequency-spatial filtering to achieve the required sub-band-specific spatial responses in Sec. \ref{sec:IV}. 
    Sec. \ref{sec:V} presents simulation results that compare the performance of Staircase TTD codebooks with state-of-the-art methods. 
    Finally, Sec. \ref{sec:VI} presents concluding remarks and future steps.

\textit{Notation:} Scalars, vectors, and matrices are denoted by non-bold, bold lower-case, and bold upper-case letters, respectively.
For a given matrix $\mathbf{A}$, $e^{\mathbf{A}}$ and $\log(\mathbf{A})$ denote matrices with the $(i,j)^{th}$ element given by $e^{A_{i,j}}$ and $\log\mathbf{A}_{i,j}$ respectively. Further, the $n^{th}$ element of a vector $\mathbf{v}$ is denoted as $\text{v}_n$. Conjugate, transpose and Hermitian transpose are denoted by $(.)^{*}$, $(.)^{\transpose}$, and $(.)^{\hermitian}$ respectively. 

%
%
\section{System Model}
\label{sec:system_model}
We consider a cellular system where a Base Station (BS) simultaneously serves $K$ users (UE) spatially distributed at angles $\theta^{(k)}$ $\forall$ $k=1,...,K$.  The BS operates over the bandwidth $BW$ and transmits an Orthogonal Frequency Division Multiplexing (OFDM) signal with a total of $M_{tot}$ subcarriers at carrier frequency $f_c$, where the frequency of the $m^{th}$ subcarrier is given by $f_m=f_c-BW/2+BW(m-1)/(M_{tot}-1)$ $\forall$ $m\in\{1,...,M_{tot}\}$. Each UE operates over a non-overlapping contiguous bandwidth $BW/K$ with a total of $M_{tot}/K$ subcarriers.

The BS is equipped with an $N_T \times 1$ analog TTD array with uniform half-wavelength spacing ($\lambda_c/2=c/(2f_c)$, where $c$ is the speed of light). Each antenna element is controlled with time delays and phase shifts, which are denoted by vectors $\boldsymbol{\tau},\boldsymbol{\Phi}\in\mathbb{R}^{N_T\times 1}$ respectively. 
The frequency-dependent precoder at the BS $\mathbf{w}_{TTD}[m]\in\mathbb{C}^{N_T\times 1}$ is thus obtained as follows:
\begin{equation}
    \mathbf{w}_{TTD}[m] = \frac{1}{\sqrt{N_T}}e^{j(2\pi f_m\boldsymbol{\tau}+\mathbf{\Phi})}
    \label{eqn:w_TTD}
\end{equation}

The goal is to design the per-antenna delays $\tau_n$ and phase shifts $\phi_n$ $\forall n\in\{1,...,N_T\}$ to generate beams with the desired sub-band to angle mapping. 
\vspace{-1mm}

\subsection{Uniform Staircase TTD codebook}
\begin{figure}[t]
    \centering
    \includegraphics[width=0.95\linewidth]{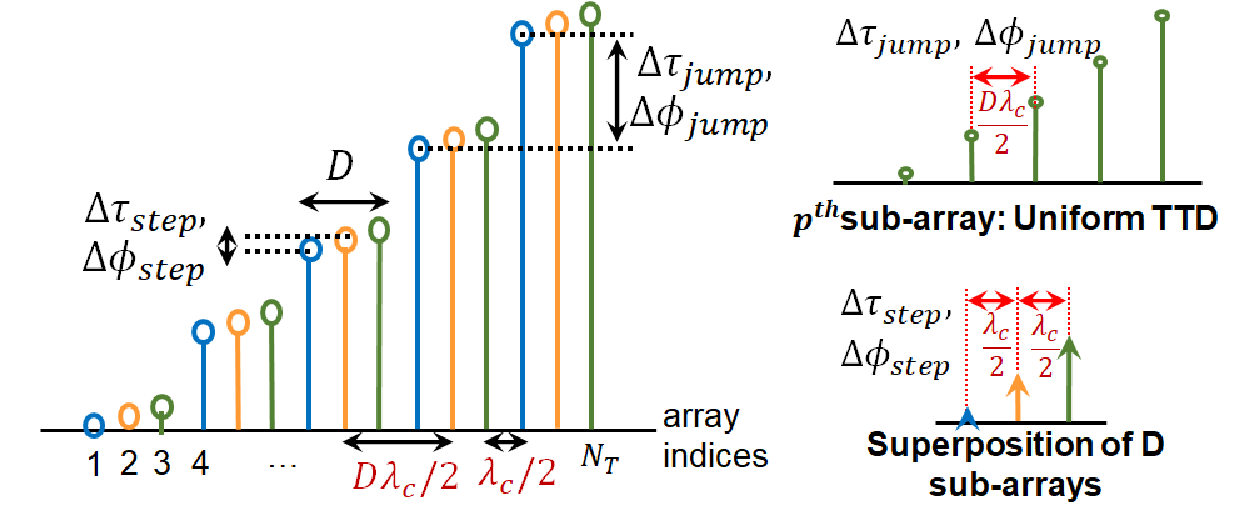}
    \caption{Uniform Staircase TTD codebook.}
    \vspace{-3mm}
    \label{fig:ramped_staircase}
\end{figure}

We introduce the uniform Staircase TTD codebook that is designed based on two sets of delay and phase increments applied at different antenna spacing intervals. The high-frequency delay and phase increments ($\Delta\tau_{h}$, $\Delta\phi_{h}$) occur at every consecutive antenna element, whereas the low-frequency increments ($\Delta\tau_{l}$, $\Delta\phi_{l}$) occur at a spacing of $D$ antenna elements. The resulting delay and phase vectors resemble a staircase function of step size $D$, where the 
delay at the $(n\text{+}1)^{th}$ antenna is given as follows:
\begin{equation}
    \begin{aligned}
        \tau_{n+1} = \left\{\begin{array}{cc}
           \tau_{n} + \Delta\tau_h + \Delta\tau_{l}  & \text{if} \hspace{2pt} \bmod{(n,D)}=0 \\
           \tau_{n} + \Delta\tau_{h}  & \text{otherwise}
        \end{array} \right.
    \end{aligned}
    \label{eqn:staircase_delays}
\end{equation}
where $\bmod(.)$ denotes the modulo operator. The per-antenna phase shifts apply increments in a similar manner. 
Under special condition $\bmod(N_T,D)=0$, it is possible to realize the Kronecker decomposition of the Staircase TTD combiner in (\ref{eqn:w_TTD}) to obtain delays and phases that can be expressed as follows: 
\begin{equation}
    \begin{aligned}
        \boldsymbol{\tau} &= \underbrace{(\Delta\tau_{l}+D\Delta\tau_{h})}_{\Delta\tau_{jump}}[0,...,\frac{N_T}{D}-1]^T \oplus [0,...,{D}-1]^T\underbrace{\Delta\tau_{h}}_{\Delta\tau_{step}}\\
        \boldsymbol{\Phi} &= \underbrace{(\Delta\phi_{l}+D\Delta\phi_{h})}_{\Delta\phi_{jump}}[0,..,\frac{N_T}{D}-1]^T \oplus [0,...,{D}-1]^T\underbrace{\Delta\phi_{h}}_{\Delta\phi_{step}}
    \end{aligned}
\end{equation}
where $\oplus$ denotes the Kronecker summation of two vectors $\mathbf{a}\in\mathbb{C}^{N_1\times 1}$ and $\mathbf{b}\in\mathbb{C}^{N_2\times 1}$, defined as $\mathbf{a}\oplus\mathbf{b}\in\mathbb{C}^{N_1 N_2\times 1}=\log (e^{\mathbf{a}}\otimes e^{\mathbf{b}})$, where $\otimes$ denotes Kronecker product. For ease of notation, we define $\Delta\tau_{jump}=\Delta\tau_{l}+D\Delta\tau_{h}$ and $\Delta\phi_{jump}=\Delta\phi_{l}+D\Delta\phi_{h}$ as the Staircase \textit{jump} parameters, and $\Delta\tau_{step}=\Delta\tau_{h}$ and $\Delta\phi_{step}=\Delta\phi_{h}$ as the \textit{step} parameters as shown in Fig. \ref{fig:ramped_staircase}, since the two parameters govern the inter- and intra- step behaviour of the staircase TTD codebook. Consequently, the delays and phases of the uniform Staircase TTD codebook in (\ref{eqn:staircase_delays}) can be expressed as follows:
\begin{equation}
    \begin{aligned}
        \tau_{n+1} = \left\{\begin{array}{cc}
            \tau_{n} +\Delta\tau_{\textit{jump}} - (D-1)\Delta\tau_{\textit{step}} ; &  \bmod{(n,D)}=0 \\
            \tau_{n} +\Delta\tau_{\textit{step}} ; & \text{otherwise}
        \end{array} \right.
    \end{aligned}
    \label{eqn:staircase_delays2}
\end{equation}
\begin{equation}
    \begin{aligned}
        \phi_{n+1} = \left\{\begin{array}{cc}
            \phi_{n} +\Delta\phi_{\textit{jump}} -(D-1)\Delta\phi_{\textit{step}}  &  \bmod{(n,D)}{\small = }0 \\
            \phi_{n} +\Delta\phi_{\textit{step}}  & \text{otherwise}
        \end{array} \right.
    \end{aligned}
    \label{eqn:staircase_phases2}
\end{equation} 

\section{Frequency-spatial analysis of Staircase TTD}
\label{sec:III}
\begin{figure*}[t]
    \centering
    \subfigure[Frequency-beam-centre map]{\includegraphics[width=0.24\linewidth]{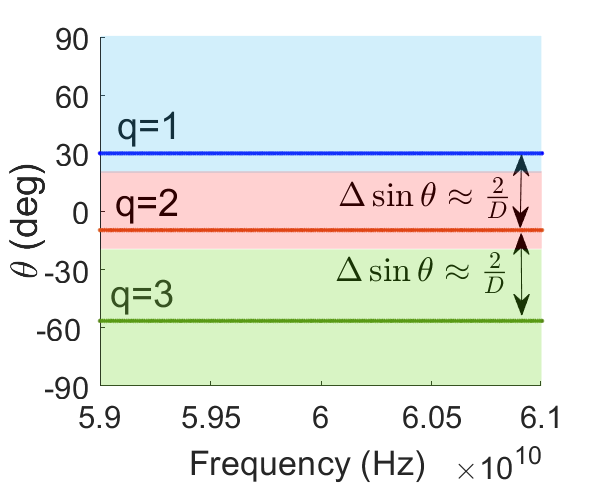}}
    \subfigure[Beamforming gain]{\includegraphics[width=0.24\linewidth]{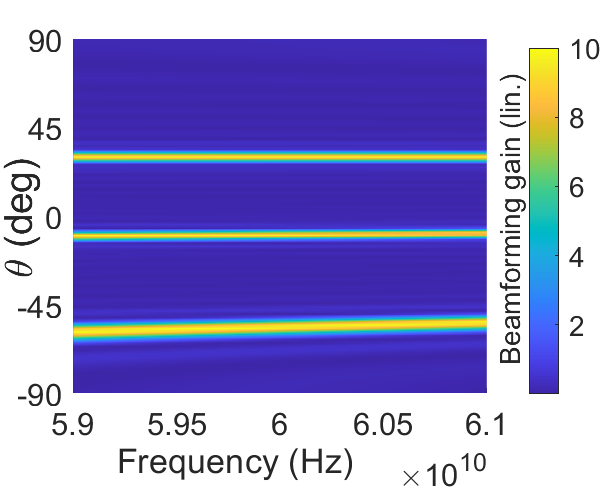}}
    \subfigure[Frequency-beam-centre map]{\includegraphics[width=0.24\linewidth]{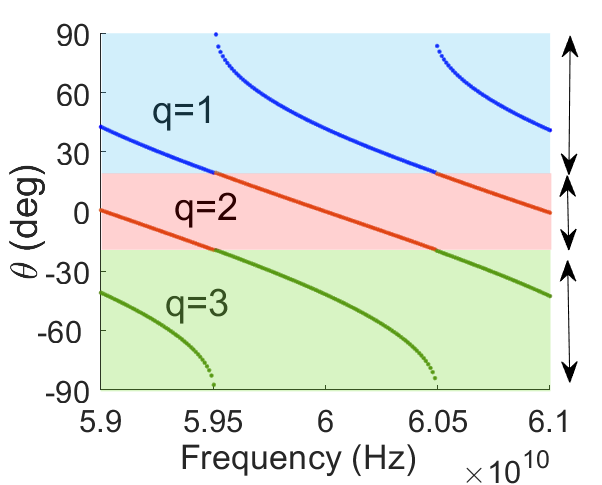}}      
    \subfigure[Beamforming gain]{\includegraphics[width=0.24\linewidth]{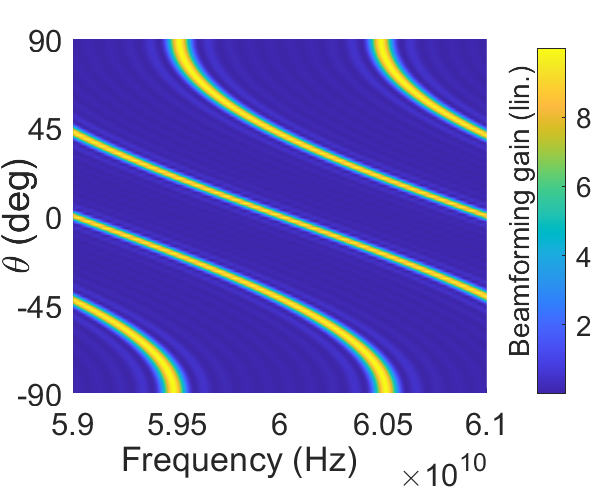}}\vspace{-2mm}
    \caption{\small Frequency-beam-centre map and beamforming gain $\tilde{G}(\theta,f_m)$ for each uniform TTD sub-array for $D=3$, $N_T/D=10$, $\Delta\phi_{jump}=0$. (\textbf{a},\textbf{b}) Directional grating lobes with $\Delta\tau_{jump}=-D\sin(\pi/6)/{2f_c}$. 
    (\textbf{c},\textbf{d}) Complete dispersion with frequency diversity, $\Delta\tau_{jump}=2/BW$. 
    }\vspace{-2.5mm}
    \label{fig:beam_mapping}
\end{figure*}

\begin{figure*}[t]
    \centering
    \subfigure[$F(\theta,f_m)$]{\includegraphics[width=0.19\linewidth]{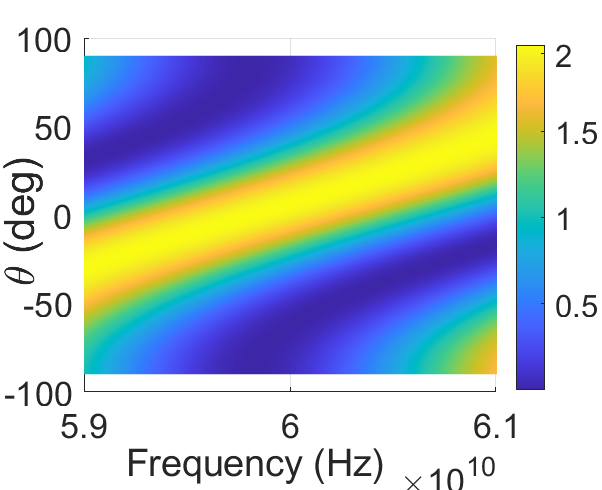}}
    \subfigure[Spatial filtering]{\includegraphics[width=0.19\linewidth]{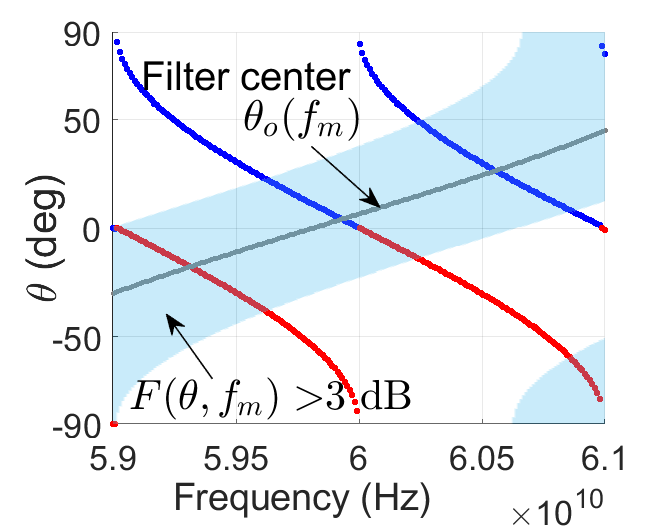}}
    \subfigure[Final gain $G(\theta,f_m)$]{\includegraphics[width=0.19\linewidth]{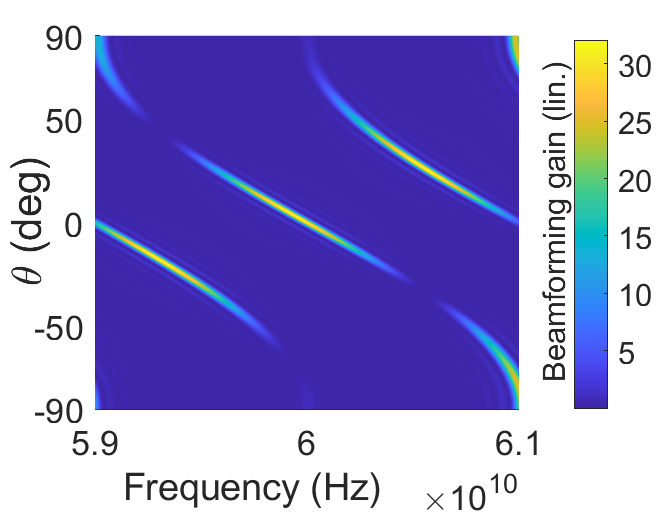}}
    \subfigure[Spatial filtering]{\includegraphics[width=0.19\linewidth]{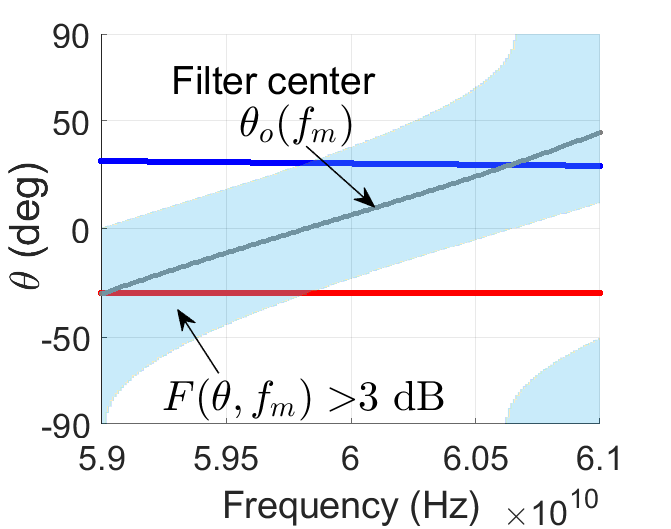}}
    \subfigure[Final gain $G(\theta,f_m)$]{\includegraphics[width=0.19\linewidth]{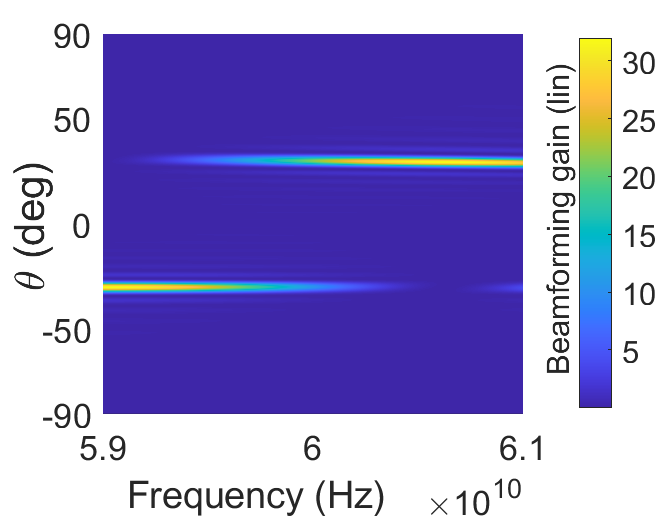}}\vspace{-2mm}
    \caption{\small Superposition of $D$ uniform TTD sub-arrays results in frequency-spatial filtering $F(\theta,f_m)$ of the parent grating lobe pattern $\tilde{G}(\theta,f_m)$. $N_T=32,D=2$, $\Delta\tau_{step}=-0.6/BW$, $\Delta\phi_{step}=0.1\pi$, $\Delta\phi_{jump}=0$. \textbf{(b,c)} $\Delta\tau_{jump}=2/BW$. \textbf{(d,e)} $\Delta\tau_{jump}=0.5/f_c$. 
    }\vspace{-3.5mm}
    \label{fig:filtering}
\end{figure*}
\subsection{Frequency-angle mapping of each sub-array}
\label{subsec:freq_angle_map}
The uniform Staircase TTD codebook can be visualized as the superposition of $D$ uniform TTD sub-arrays (shown in Fig. \ref{fig:ramped_staircase}) with antenna spacing $D\lambda_c/2$, delay spacing $\Delta\tau_{jump}$ and phase spacing $\Delta\phi_{jump}$. Since the antenna spacing exceeds the critical $\lambda_c/2$ spacing by a factor of $D$, the resulting beams exhibit $D$ grating lobes or spectral copies for each frequency. 

Each sub-array would have an identical frequency-beam-centre mapping owing to identical uniform TTD array parameters. Based on (\ref{eqn:w_TTD}), the precoder for each sub-array $\mathbf{\tilde{w}}_{TTD}[m]\in\mathbb{C}^{N_T/D\times 1}$ is determined by: 
\begin{equation}
    \mathbf{\tilde{w}}_{TTD}[m] = {\sqrt{\frac{D}{N_T}}}e^{j\pi[0,...,\frac{N_T}{D}-1]^T(2 f_m \Delta\tau_{jump} + \Delta\phi_{jump}/\pi)} 
    \label{eqn:w_ttd_sub}
\end{equation}
The array response vector $\mathbf{\tilde{a}}_{D}(\theta,f_m)\in\mathbb{C}^{N_T/D\times 1}$ for each sub-array with $D\lambda_c/2$ antenna-spacing at an angle of arrival $\theta$ can be given as follows:
\begin{equation}
    \mathbf{\tilde{a}}_{D}(\theta,f_m)=e^{-j\pi \frac{f_m}{f_c} [0,...,\frac{N_T}{D}-1]^TD\sin\theta} 
    \label{eqn:array_response}
\end{equation}

The frequency-dependent beamforming gain at angle $\theta$ can thus be obtained as $\tilde{G}(\theta,f_m)=|\mathbf{\tilde{w}}_{TTD}^H[m]\mathbf{\tilde{a}}_D(\theta,f_m)|^2$, which can be simplified as follows:
\begin{equation}
    \tilde{G}(\theta,f_m)=\left|\frac{\sin\left(\frac{N_T}{D}\frac{\pi}{2} \Psi_{jump}(f_m)\right)}{\sin\left(\frac{\pi}{2} \Psi_{jump}(f_m)\right)}  \right|^2
    \label{eqn:gain_jump}
\end{equation}
where $\Psi_{jump}(f_m)=2f_m\Delta\tau_{jump}+\Delta\phi_{jump}/\pi+D(f_m/f_c) \sin\theta$. The beam-centre for frequency $f_m$, denoted by $\theta^\star(f_m)$ or $\theta_m^\star$, corresponds to the angle that maximizes the beamforming gain function, i.e. $\theta_m^\star = \{\theta|G(\theta,f_m)=N_T/D\}$, and can be obtained by solving $\Psi_{jump}(f_m)=2z$, $z\in\mathbb{Z}$. Owing to grating lobes, each frequency $f_m$ will have $D$ beam-centre solutions, which are given as follows:
\begin{equation}
\begin{aligned}
    \theta^\star(f_m&,q) = \sin^{-1}\Big[1-\frac{2}{D}(q-1)\frac{f_c}{f_m} -\\
    &mod\left(2f_c\frac{\Delta\tau_{jump}}{D}+\frac{\Delta\phi_{jump}}{D\pi}\frac{f_c}{f_m} +1, 2\frac{f_c}{D f_m} \right)  \Big] 
\end{aligned}    
\label{eqn:beam_centres}
\end{equation}
where each value of $q=1,...,D$ corresponds to a distinct spectral copy of the main beam. As is evident from (\ref{eqn:beam_centres}), the $D$ spectral copies for each frequency $f_m$ have an angular separation of $\Delta\sin\theta_m^\star = \frac{2}{D}\frac{f_c}{f_m}\approx \frac{2}{D}$ when $f_c>>BW$. 
Thus, the $D\lambda_c/2$ array-spacing partitions the angular region into $D$ non-overlapping segments of uniform sinusoidal width, within which each spectral copy is confined, as shown in Fig. \ref{fig:beam_mapping}(a,c). The grating factor $D$ thus determines the number and relative spacing of spectral beam copies. 

Further, the slope of the frequency-beam-centre map, denoted by $\frac{\partial \sin\theta_m^\star}{\partial f_m}$ can be obtained from (\ref{eqn:beam_centres}) as $-\frac{2\Delta\tau_{jump}}{D}$. This tells us that $\Delta\tau_{jump}$ determines the extent of frequency-dependent angular dispersion of each spectral copy within its segment. 
Setting $\Delta\tau_{jump}=-\frac{D\sin\theta_o}{2f_c}$ creates a directional beam at $\theta_o$ $\forall f_m$, with spectral copies at $\theta_m^\star=\sin^{-1}(\bmod(\sin\theta_o-2\frac{q-1}{D}\frac{f_c}{f_m}+1,2)-1)|_{ q=2,...,D}$, as seen in Fig. \ref{fig:beam_mapping}(a,b). When $\frac{1}{fc}<<|\Delta\tau_{jump}|<\frac{1}{BW}$, each spectral copy exhibits partial dispersion within its respective spectral segment. When $|\Delta\tau_{jump}|\geq\frac{1}{BW}$, each spectral copy maps to its entire angular segment in at least one mapping cycle, as seen in Fig. \ref{fig:beam_mapping}(c,d). 

\subsection{Superposition of the $D$ sub-arrays: Spatial filtering}
Section \ref{subsec:freq_angle_map} obtains the beamforming gain $\tilde{G}(\theta,f_m)$ (\ref{eqn:gain_jump}) and frequency-angle mapping $\theta_m^\star$ of grating lobes (\ref{eqn:beam_centres}) for the $D$ identical uniform TTD sub-arrays that constitute the Staircase TTD codebook. 
Since these $D$ sub-arrays are uniformly separated in space ($\lambda_c/2$ antenna spacing), time ($\Delta\tau_{step}$) and phase ($\Delta\phi_{step}$), as shown in Fig. \ref{fig:ramped_staircase}, the effective phase separation between adjacent sub-arrays can be expressed as $\pi\Psi_\mathrm{o}(f_m)$, where $\Psi_\mathrm{o}(f_m)=2 f_m\Delta\tau_{step} + (f_m/f_c) \sin\theta+\Delta\phi_{step}/\pi$. Thus, the overall beamforming gain $G(\theta,f_m)$ of the entire Staircase TTD codebook can be expressed as the exponentially weighted sum of $\tilde{G}(\theta,f_m)$, as shown in (\ref{eqn:overall_gain0}), which can be simplified to obtain (\ref{eqn:overall_gain}):
\begin{equation}
    \begin{aligned}
         G(\theta,f_m) &= \big|\sum_{q=1}^D e^{-j\pi (q-1) \Psi_\mathrm{o}(f_m)} . \mathbf{\tilde{w}}_{TTD}^H[m]\mathbf{\tilde{a}}_{(D)}\big|^2        
    \end{aligned}
    \label{eqn:overall_gain0}
\end{equation}
\begin{equation}
    \begin{aligned}         
        G(\theta,f_m) &=\tilde{G}(\theta,f_m) ~.~ \underbrace{\Big|\frac{\sin\left(({D\pi}/{2}) \Psi_o(f_m)\right)}{\sin\left(({\pi}/{2}) \Psi_o(f_m)\right)} \Big|^2}_{\text{Spatial filter: } F(\theta,f_m)}
    \end{aligned}
    \label{eqn:overall_gain}
\end{equation} 
 The term $F(\theta,f_m)=\big|\frac{\sin\left(({D\pi}/{2}) \Psi_o(f_m)\right)}{\sin\left(({\pi}/{2}) \Psi_o(f_m)\right)} \big|^2$ represents the frequency-spatial filter response that results from the superposition of the $D$ TTD sub-arrays, uniformly separated in phase, space and time. 
 The filter $F(\theta,f_m)$ is centred at angle $\theta_\mathrm{o}(f_m)$, which corresponds to the gain maximizing trajectory about which the filter's spatial response is symmetric, and can be obtained by solving $\Psi_\mathrm{o}(f_m)=2z$, $z\in\mathbb{Z}$, as follows:
 \begin{equation}
 \small
   \theta_\mathrm{o}(f_m)=\sin^{-1}\left(1-\bmod(2f_c\Delta\tau_{step}+\frac{\Delta\phi_{step}}{\pi}\frac{f_c}{f_m} +1, 2\frac{f_c}{f_m}) \right)
    \label{eqn:rainbow_filter}
\end{equation}

The \textit{step} delay $\Delta\tau_{step}$ makes the filter's spatial response frequency-dependent as seen in Fig. \ref{fig:filtering}(a). This is reminiscent of dispersive rainbow beam codebooks constructed using uniform TTD arrays in \cite{vb_ttd_2021,yan_ttd_2019,9154233,3D_rainbow}. 
Further, the 3dB angular width of the filter for a given $f_m$ is given by  $\Delta\sin\theta=\frac{2\times 0.886}{D}$ \cite[Chapt 22.7]{array_textbook1}. 
Thus, for each frequency, the filter retains beam patterns corresponding to roughly one spectral segment of angular width $\Delta\sin\theta \approx\frac{2}{D}$ out of the $D$ spectral copies present in the parent beam-pattern $\tilde{G}(\theta,f_m)$ as shown in Fig. \ref{fig:filtering}(b,d), thereby resulting in the sub-band-specific spatial responses shown in Fig. \ref{fig:filtering}(c,e). Through the systematic design of grating lobe parameters ($D$, $\Delta\tau_{jump}$, $\Delta\phi_{jump}$) and filter parameters ($\Delta\tau_{step}$, $\Delta\phi_{step}$), we can achieve the required directional sub-band-specific beams.

\section{Two-stage design of sub-band-beams}
\label{sec:IV}
In this section, we propose the two-stage design of the Staircase codebook parameters $\Delta\tau_{jump}$, $\Delta\phi_{jump}$, $\Delta\tau_{step}$, $\Delta\phi_{step}$ and $D$ defined in (\ref{eqn:staircase_delays2}) and (\ref{eqn:staircase_phases2}), to construct 
sub-band-specific beams to simultaneously communicate with $K$ users located at sinusoidally equidistant angles $\theta^{(q)}$ $\forall q\in\{1,...,K\}$ in the sector $[\theta_1,\theta_2]$,
with uniform $(BW/K)$ sub-band assignment to each user, as shown in Fig. \ref{fig:desired_beams}(a). The $K$ UE angles $\theta^{(q)}$ $\forall q\in\{1,...,K\}$ are given as follows:
\begin{equation}
    \theta^{(q)}=\sin^{-1}\left(\sin\theta_1+(q-1)\frac{\sin\theta_2-\sin\theta_1}{K-1}\right) \hspace{2pt} 
    \label{eqn:target_angles}
\end{equation} 

\subsection{Sub-band beam design with uniform Staircase codebooks}
\label{subsec:design_A}
\begin{figure}[t]
    \centering
    \includegraphics[width=\linewidth]{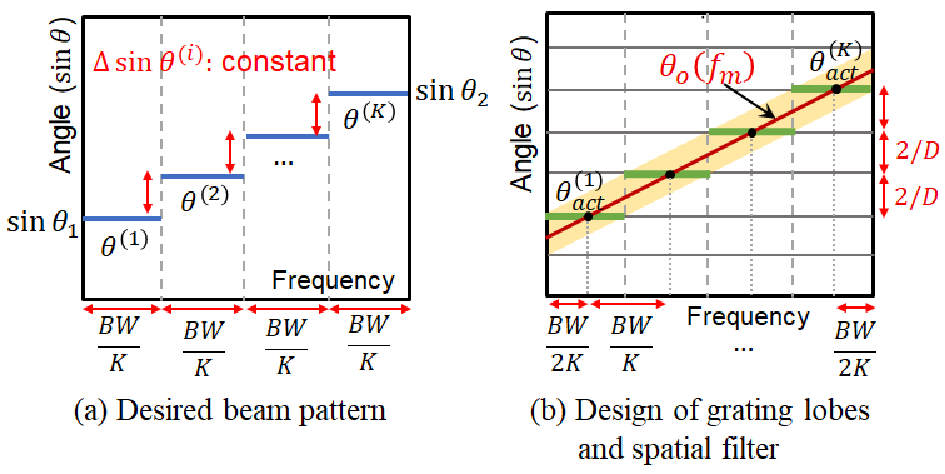}\vspace{-3mm}
    \caption{\small (a) Target sub-band-angle mapping. (b) Design of grating lobes and spatial filter $F(\theta,f_m)$ to achieve the beam-pattern in (a).}\vspace{-3mm}
    \label{fig:desired_beams}    
\end{figure}
\textbf{Stage I:} The first step towards designing the required beam pattern is constructing $K$ directional grating lobes exactly at the required angles $\theta^{(q)}$ $\forall q\in\{1,...,K\}$ in (\ref{eqn:target_angles}), as shown in Fig. \ref{fig:desired_beams}(b). We know that the angular separation between adjacent grating lobes is $\frac{2}{D}\frac{f_c}{f_m}$, where $D\in\mathbb{Z}$ is the step size of the uniform Staircase codebook. Hence, in order to fit exactly $K$ grating lobes in $[\theta_1,\theta_2]$, we must select $D$ as the smallest integer satisfying $\gamma|\sin\theta_2-\sin\theta_1| \geq (K-1)\frac{2}{D}$, where $\gamma=1+\frac{BW}{2f_c}-\frac{BW}{2Kf_c}$ is the beam-squint\footnote{Upon setting $\Delta\tau_{jump}=-\frac{D\sin\theta_1}{2f_c}$, all spectral copies except the first copy at $\theta_1$, exhibit beam-squint. Hence, the angular separation between the first and $K^{th}$ grating lobes is $\frac{2(K-1)}{\gamma D}$ where $\gamma=\frac{1}{f_c}\left(f_c+\frac{BW}{2}-\frac{BW}{2K}\right)$} 
correction factor. Thus, $D$ can be computed as follows: 
\begin{equation}
\begin{aligned}
    D = \Bigg\lceil \frac{2(K-1)}{\gamma|\sin\theta_2-\sin\theta_1|} \Bigg\rceil
\end{aligned}    
\label{eqn:D_req}
\end{equation}
Further, setting $\Delta\tau_{jump}=\frac{-D\sin\theta_1}{2f_c}$ and $\Delta\phi_{jump}=0$ creates $D$ grating lobes at $\theta_{act}^{(i)}$ $\forall i=1,...,D$, given as follows, out of which $\theta_{act}^{(q)}|_{ q=1,...,K}$ fall in the range $[\theta_1,\theta_2]$.
\begin{equation}
    \theta_{act}^{(q)}=\sin^{-1}\left(\bmod\left(\sin\theta_1+(q-1)\frac{2}{D}+1,2\right)-1\right) \hspace{2pt} 
    \label{eqn:actual_angles}
\end{equation}

\begin{figure}[t]
    \centering        
    \subfigure[$\Delta\tau_{step}=-0.58/BW,\Delta\phi_{step}=0.46\pi$]{\includegraphics[width=0.92\linewidth]{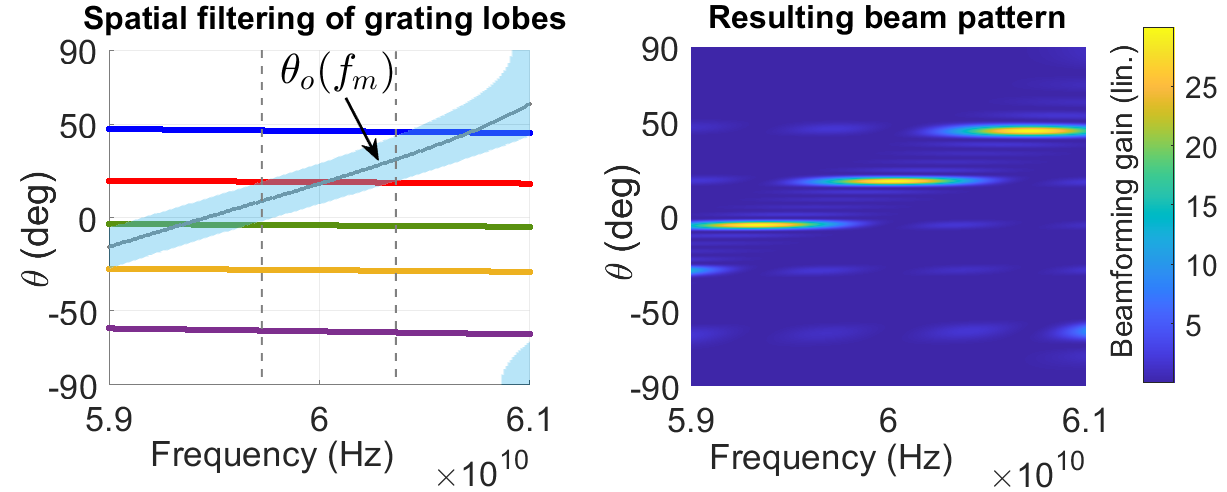}}
    \subfigure[$\Delta\tau_{step}=1/{BW},~\Delta\phi_{step}=0.075\pi$]{\includegraphics[width=0.92\linewidth]{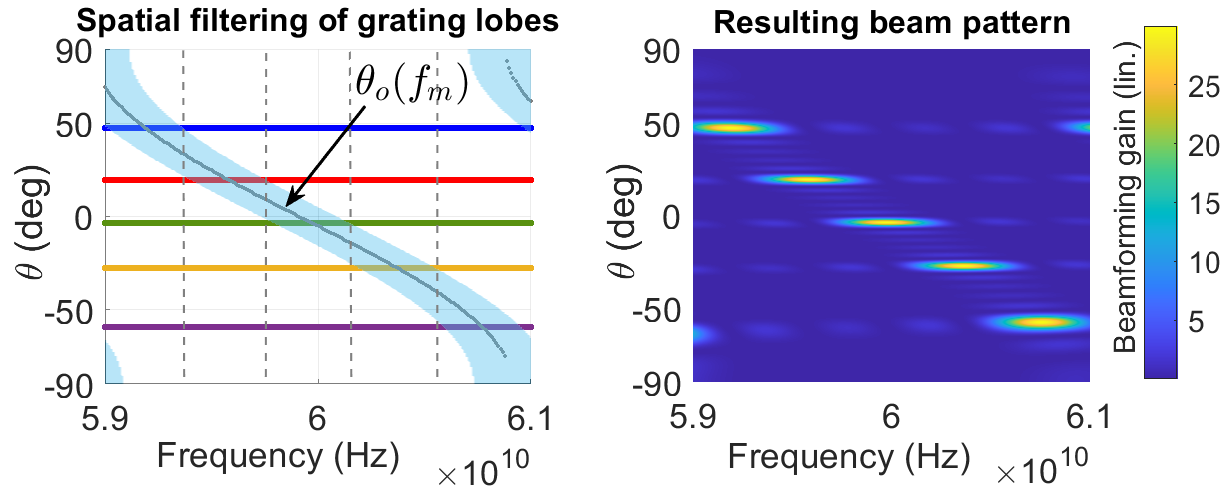}}
    \subfigure[$\Delta\tau_{step}=-1.45/BW,\Delta\phi_{step}=0$]{\includegraphics[width=0.92\linewidth]{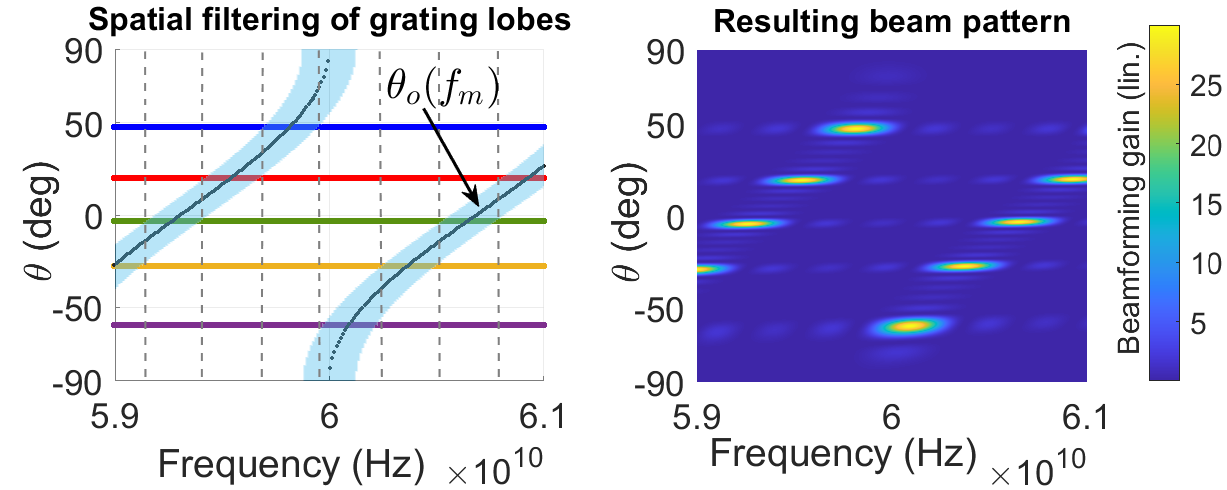}}
    \vspace{-2.5mm}
    \caption{\small Effect of spatial filter parameters $\{\Delta\tau_{step}$,$\Delta\phi_{step}\}$ on sub-band-beam patterns for $D=5,\Delta\tau_{jump}=2.16/f_c,\Delta\phi_{jump}=0$.}
    \vspace{-4.5mm}
    \label{fig:filter_design_2}
\end{figure}

\textbf{Stage II:} The next step is to design the frequency-spatial filter $F(\theta,f_m)$ to achieve the desired sub-band-specific filtering of the grating lobes as shown in Fig. \ref{fig:desired_beams}(b). For given grating lobes at $\theta_{act}^{(i)}|_{i=1,...,D}$, the choice of filter parameters $\Delta\tau_{step}$ and $\Delta\phi_{step}$ determines the exact sub-band-angle mapping achieved, as is seen in the examples in Fig. \ref{fig:filter_design_2}.  
In order to ensure $K$ equal sub-bands that map to the $K$ angles $\theta_{act}^{(q)}|_{ q=1,...,K}$, we need to design  $\Delta\tau_{step}$ and $\Delta\phi_{step}$ in a manner as to make the filter-centre trajectory $\theta_\mathrm{o}(f_m)$ intersect the $K$ grating lobes at the centres of the respective sub-bands, as shown in Fig. \ref{fig:desired_beams}(b). 

For example, to achieve the beam pattern in Fig. \ref{fig:desired_beams}(a), the first sub-band centred at $f^{(1)}=f_c-BW/2+BW/(2K)$ must map to $\theta_{act}^{(1)}=\theta_1$ whereas the $K^{th}$ sub-band centred at $f^{(K)}=f_c+BW/2-BW/(2K)$ must map to $\theta_{act}^{(K)}=\theta_2$. 
Consequently, $\Delta\tau_{step}$ and $\Delta\phi_{step}$ \footnote{$\Delta\phi_{step}$ is obtained by solving $\theta_\mathrm{o}(f^{(K)})=\theta_2$ in (\ref{eqn:rainbow_filter}) with a substitution of $\Delta\tau_{step}$ from (\ref{eqn:dtau_step_der}), which upon simplification gives (\ref{eqn:dphi_step_der}).} 
can be obtained as follows: 
\begin{equation}
    \Delta\tau_{step}=-\frac{1}{2}\frac{\partial\sin\theta_\mathrm{o}(f_m)}{\partial f_m}=\frac{{f^{(1)}}\sin{\theta}_1-{f^{(K)}}\sin{\theta}_2}{2f_c(K-1)\frac{BW}{K}}
    \label{eqn:dtau_step_der}
\end{equation}
\begin{equation}
\begin{aligned}
    \Delta\phi_{step}=-\pi\frac{f^{(K)}}{f_c}\left(\sin\theta_2 + 2f_c\Delta\tau_{step}\right)
\end{aligned}    
\label{eqn:dphi_step_der}
\end{equation}\vspace{-3mm}
\subsection{Mapping discrepancies with uniform Staircase codebooks}
\begin{figure}[t]
    \centering
    \subfigure[$K=3, D=4$]{\includegraphics[width=0.48\linewidth]{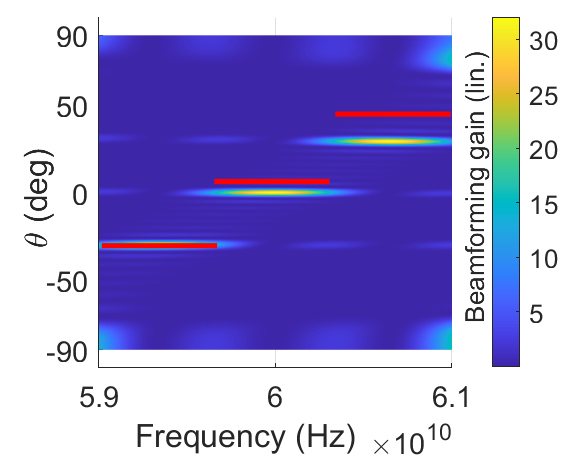}}
    \subfigure[$K=3, D=3.31$]{\includegraphics[width=0.48\linewidth]{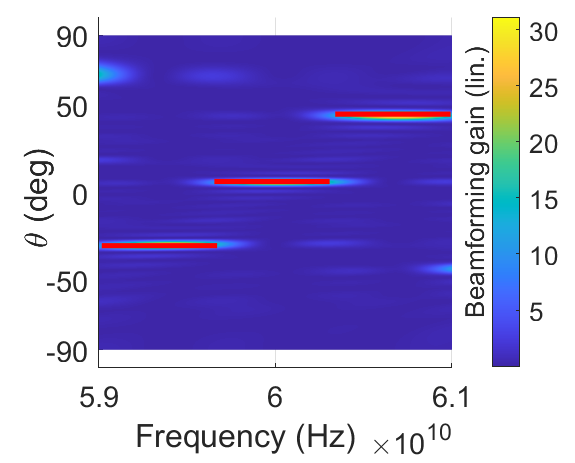}}
    \vspace{-1.5mm}
    \caption{\small \textbf{(a)} Uniform Staircase codebook (\ref{eqn:staircase_delays2}),(\ref{eqn:staircase_phases2}) enforces $D\in\mathbb{Z}$, resulting in discrepancy between target (shown in red) and actual sub-band-angle maps. \textbf{(b)} New Staircase (\ref{eqn:new_staircase}) allows $D\in\mathbb{R}$, thereby resolving mapping discrepancy. Here, $\{\theta_1,\theta_2\}=\{-\pi/6,\pi/4\}$.}
    \label{fig:mapping_discrepancy}\vspace{-4mm}
\end{figure}
The first step to generating directional sub-band-specific beams mapped to angles $\theta^{(q)}\big|_{q=1,...,K}$ as shown in (\ref{eqn:target_angles}), involves setting $\Delta\tau_{jump}=-\frac{D\sin\theta_1}{2f_c}$ and $\Delta\phi_{jump}=0$. 
This results in grating lobes at angles $\theta_{act}^{(i)}|_{ i=1,...,K}$ as shown in (\ref{eqn:actual_angles}). Since the uniform Staircase codebook constrains the (uniform) step-size $D$ to be an integer, designing $D$ as per (\ref{eqn:D_req}) results in a mismatch or discrepancy between the target and actual angular levels, i.e. $\theta^{(q)}\neq \theta_{act}^{(q)}$ $\forall q\in\{2,...,K\}$, as can be seen in Fig. \ref{fig:mapping_discrepancy}(a), where the target sub-band-angle map is shown in red. This can be verified by substituting $D=\lceil\frac{2(K-1)}{\gamma|\sin\theta_2-\sin\theta_1|} \rceil$ into (\ref{eqn:actual_angles}) and comparing with (\ref{eqn:target_angles}). 
Thus, staircase TTD codebooks with uniform step-size suffer from mapping discrepancies which inhibit our ability to achieve the desired sub-band-angle map.  

\vspace{-1.0mm}
\subsection{Alternative Staircase to overcome mapping discrepancies}
\label{subsec:alternative}
In this section, we formulate a Staircase TTD codebook with non-uniform step-size, relaxing the requirement of $D$ being an integer. The uniform Staircase TTD codebook described in (\ref{eqn:staircase_delays2}) and (\ref{eqn:staircase_phases2}) can be visualized as having element-wise increments of $\Delta\tau_{step}$ with \textit{wrapping around} by a magnitude of $-(\Delta\tau_{jump}-D\Delta\tau_{step})$ occurring at every $n^{th}$ array element satisfying $\bmod(n-1,D)=0$. This \textit{wrapping around} is triggered by the array index $n$ and results in a Staircase codebook with uniform integer step-size $D$. 

Instead, we can define a new Staircase TTD codebook where the wrapping around is triggered every time a certain magnitude threshold is exceeded, in the following manner.
\begin{equation}
    \begin{aligned}
        \tau_n&=\bmod\left((n-1)\Delta\tau_{step},D\Delta\tau_{step}-\Delta\tau_{jump}\right)\\
        \phi_n&=\bmod\left((n-1)\Delta\phi_{step},D\Delta\phi_{step}-\Delta\phi_{jump}\right)
    \end{aligned}
    \label{eqn:new_staircase}
\end{equation}
This new formulation results in a Staircase TTD codebook with non-uniform step-size. Thus, the parameter $D$, which now controls only the angular spacing between grating lobes, is no longer constrained to be an integer, and can be selected as:
\begin{equation}
    D=\frac{2(K-1)}{\gamma\left(\sin\theta_2-\sin\theta_1\right)}
    \label{eqn:new_D}
\end{equation}
With $\Delta\tau_{jump}=-\frac{D\sin\theta_1}{2f_c}$, the actual grating lobes now coincide with the target angular levels $\theta^{(q)}=\theta_{act}^{(q)}\forall$ $q=1,...,K$, thereby resolving the mapping discrepancy as seen in Fig. \ref{fig:mapping_discrepancy}(b). 

Table \ref{tab:Staircase_TTD_params} summarises the Staircase TTD codebook design to achieve sub-band-specific beams shown in Fig. \ref{fig:desired_beams}(a).  
\begin{figure}[t]
    \centering
    {\includegraphics[width=\linewidth]{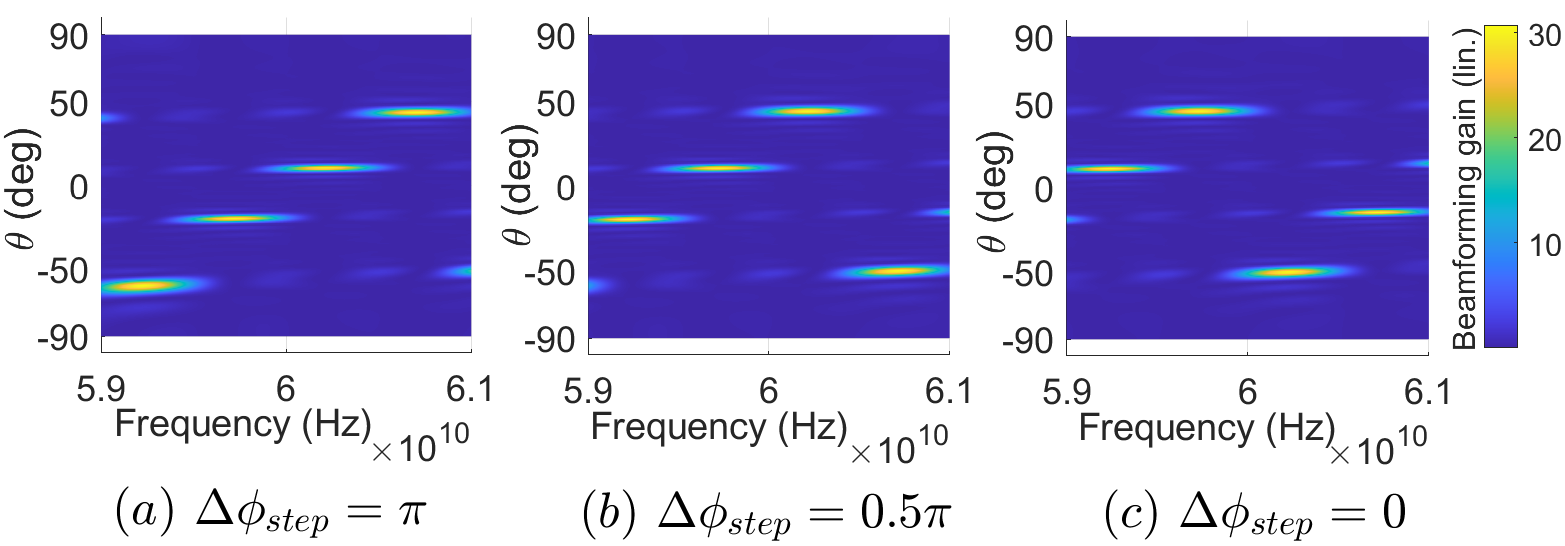}}\vspace{-1.5mm}
    \caption{\small Cyclical rotation of sub-band-angle mapping by changing $\Delta\phi_{step}$. $\Delta\tau_{jump}=\frac{1.63}{f_c}$, $\Delta\tau_{step}=-\frac{1.05}{BW}$, $D=3.77$.}\vspace{-3mm}
    \label{fig:rotation}
\end{figure}

\subsection{Constraints on achievable sub-band-angle mappings}
\begin{figure*}[t]
    \centering
    \subfigure[Spectral efficiency vs $K$]{\includegraphics[width=0.45\linewidth]{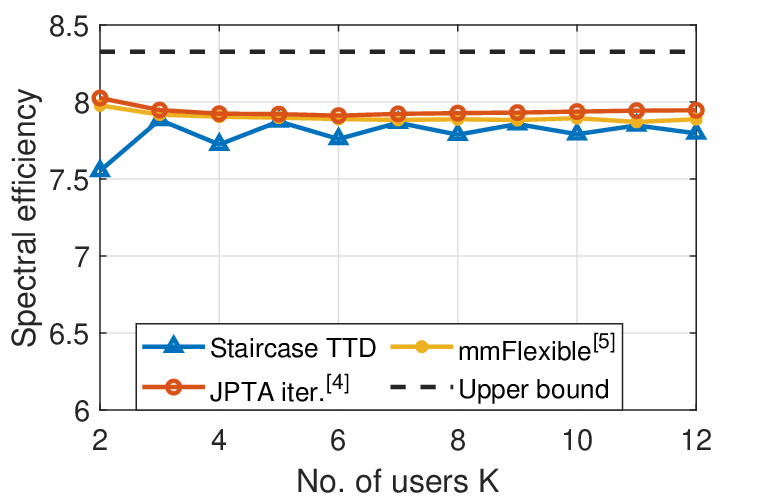}\label{fig:speff_vs_K}}  
    \subfigure[Spectral efficiency vs $BW$]{\includegraphics[width=0.45\linewidth]{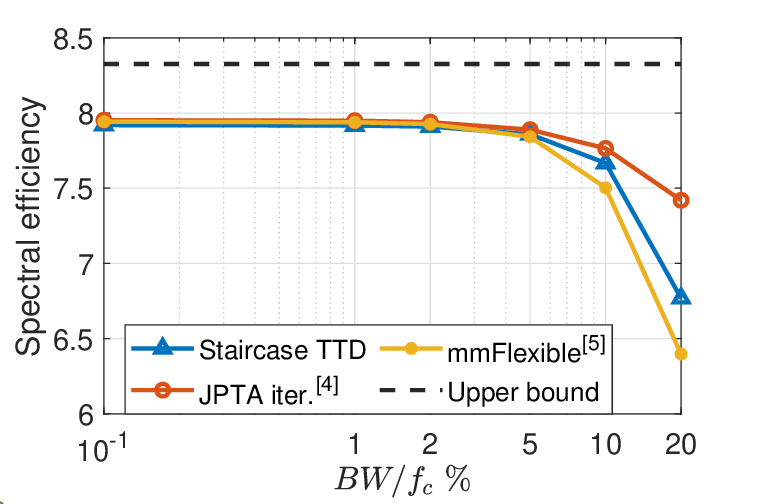}\label{fig:speff_vs_BW}} 
    \subfigure[Spectral efficiency vs $N_T$]{\includegraphics[width=0.45\linewidth]{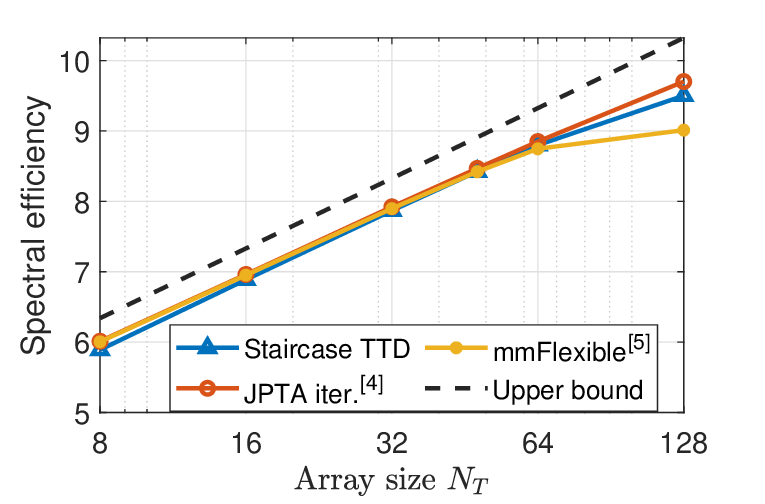}\label{fig:speff_vs_N}}
    \subfigure[Spectral efficiency vs SNR]{\includegraphics[width=0.45\linewidth]{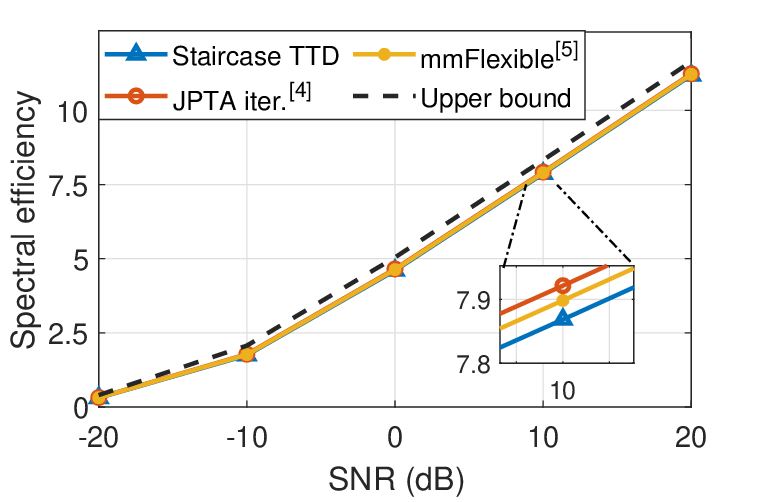}\label{fig:speff_vs_SNR}}        
    
    \vspace{-2.5mm}
    \caption{\small Performance evaluation of Staircase TTD sub-band-specific beams for multi-user data communication. Here, $f_c=60 GHz$, $BW=2GHz$, $K=5$, $N_T=32$, $\theta_1,\theta_2\in(-75^\circ,75^\circ)$, and $SNR=10dB$ unless specified otherwise.}\vspace{-4mm}
    \label{fig:my_label}
\end{figure*}

\begin{figure}[t]
    \centering
    \subfigure[Beam gain: $K=2$, $\{\theta_1,\theta_2\}=\{-30^\circ,40^\circ\}$]{\includegraphics[width=\linewidth]{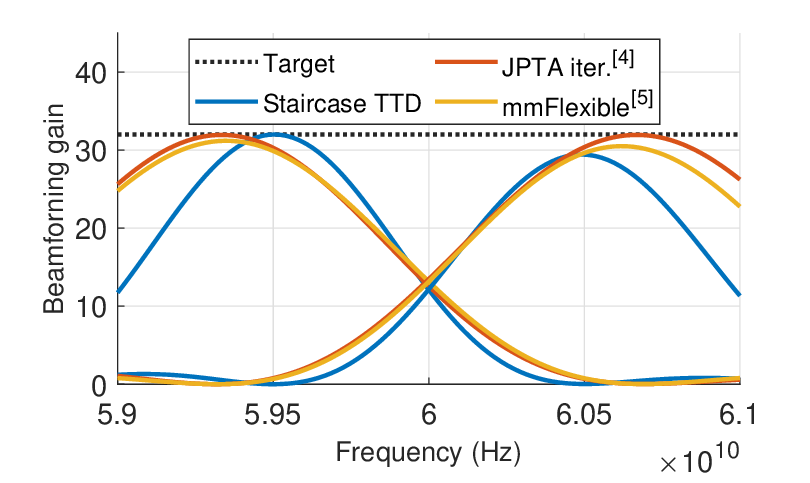}\label{fig:beam_gain_K2}}   
    \subfigure[Beam gain: $K=5$, $\{\theta_1,\theta_2\}=\{-30^\circ,40^\circ\}$]{\includegraphics[width=\linewidth]{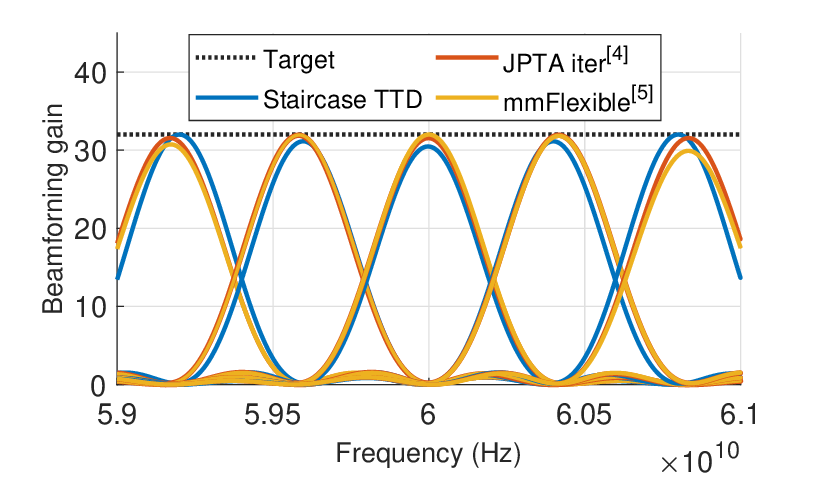}\label{fig:beam_gain_K5}}     
    
    \vspace{-1mm}
    \caption{\small On-target beamforming gain $\mathcal{B}_k(f_m)$ (eqn. (\ref{eqn:target_gain})): achieved gain at target angles $\theta^{(k)}$ $\forall ~{k=1,...,K}$, with $\{\theta_1,\theta_2\}=\{-30^\circ,40^\circ\}$, $f_c=60 GHz$, $BW=2GHz$, and $N_T=32$.}\vspace{-3.5mm}
    \label{fig:my_label}
\end{figure}
The Staircase TTD codebook formulation described in Sec. \ref{subsec:alternative} can realize sub-band-beams that map to sinusoidally equidistant angles (\ref{eqn:target_angles}) in a specified sector $[\theta_1,\theta_2]$ in monotonically increasing ($\theta_1<\theta_2$) or monotonically decreasing ($\theta_1>\theta_2$) patterns. For a given array size $N_T$, a sub-band angle map occupying the sector $[\theta_1,\theta_2]$ can be realized only if the following condition, which ensures that the array is large enough to induce \textit{wrapping around}, holds:
\begin{equation}
    \Big\lceil\frac{2(K-1)}{\gamma\left(\sin\theta_2-\sin\theta_1\right)}\Big\rceil<N_T
    \label{eqn:feasible_patterns}
\end{equation}
Further, cyclic rotations of the monotonic sub-band-angle maps as shown in Fig. \ref{fig:rotation} are possible only when $\gamma|\sin\theta_2-\sin\theta_1|>2({K-1})/({K+1})$, and can be achieved by merely changing the filter parameter $\Delta\phi_{step}$, keeping all other codebook parameters fixed. For example, we can map the first sub-band centred at $f^{(1)}=f_c-BW/2+BW/(2K)$ to angle $\theta^{(i)}$, $i\in\{1,...,K\}$ by setting $\Delta\phi_{step}$ as $\Delta\phi_{step}=-\pi\left(\frac{f^{(1)}}{f_c}\sin\theta^{(i)}+2f^{(1)}\Delta\tau_{step}\right)$.
\begin{table}[!h]
    \small
    \doublespacing
    \begin{tabular}{|m{0.4cm} m{7.6cm}|}
    \hline 
    \multicolumn{2}{|l|}{\textbf{Given:} $K$ UE at angles $\theta^{(q)}|_{q=1,...,K}\in[\theta_1,\theta_2]$, $\theta_1\neq\theta_2$}\\
    \multicolumn{2}{|l|}{
    BS has $N_T\times 1$ Analog TTD array. $\gamma=1+\frac{BW}{2f_c}-\frac{BW}{2Kf_c}$} \\
    \hline
    \multicolumn{2}{|l|}{Design TTD delays and phase shifts $\boldsymbol{\tau},\boldsymbol{\Phi}\in\mathbb{R}^{N_T\times 1}$ as follows:}\\
    1. & $D=\frac{2(K-1)}{\gamma\left(\sin\theta_2-\sin\theta_1\right)}$; $\Delta\tau_{jump}=-\frac{D\sin\theta_1}{2f_c}$; $\Delta\phi_{jump}=0$\\
    2. &  $\Delta\tau_{step}$, $\Delta\phi_{step}$ based on (\ref{eqn:dtau_step_der}) and (\ref{eqn:dphi_step_der}).\\
    3. & $\begin{aligned}
        \tau_n&=\bmod\left((n-1)\Delta\tau_{step},D\Delta\tau_{step}-\Delta\tau_{jump}\right)\\
        \phi_n&=\bmod\left((n-1)\Delta\phi_{step},D\Delta\phi_{step}-\Delta\phi_{jump}\right)
    \end{aligned}$ \\    
    \hline    
     \multicolumn{2}{|l|}{\hspace{8pt}$\theta^{(q)}=\theta_{act}^{(q)}=\sin^{-1}\left(\sin\theta_1+(q-1)\frac{\sin\theta_2-\sin\theta_1}{K-1}\right)\big|_{q=1,...,K}$ }\\
    \hline
    \end{tabular}
    \caption{Staircase TTD codebook design to realize sub-band-specific beams described in Sec. \ref{sec:system_model} and shown in Fig. \ref{fig:desired_beams}.}
    \label{tab:Staircase_TTD_params}
\end{table}

\section{Numerical Results}
\label{sec:V}
This section studies the performance of sub-band-beams designed using the Staircase TTD codebook for the system model described in Sec. \ref{sec:system_model}, in terms of the spectral efficiency of the $1$ BS and $K$ UE network. 
We present performance comparison with state-of-the-art methods, namely, the iterative weighted Least Squares optimization algorithm (JPTA iter.) presented in \cite{Samsung_1} with 20 training iterations, and the closed-form Least Squares solution (mmFlexible) proposed by \cite{UCSD}. All methods are compared with the theoretical upper bound represented by the ideal best-case beam. The BS operates at $f_c=60GHz$ with $M_{tot}=4096$ subcarriers. We consider $BW=2GHz$, $N_T=32$, $K=5$, and Signal-to-Noise Ratio (SNR) of $10 dB$, unless specified otherwise. Spectral efficiency results are averaged over all realizable beam patterns as per Table \ref{tab:Staircase_TTD_params} for $\{\theta_1,\theta_2\}\in[-75^\circ,75^\circ]$.



Fig. \ref{fig:speff_vs_K} studies spectral efficiency as a function of the number of users $K$ (or sub-bands), for $\{\theta_1,\theta_2\}\in[-75^\circ,75^\circ]$. JPTA iter\cite{Samsung_1} performs the best for all $K$, closely followed by mmFlexible\cite{UCSD}. For $K=2$, Staircase TTD suffers noticeable degradation compared to JPTA iter and mmFlexible. However, as the number of users increases ($K>4$), the performance of Staircase TTD matches up to that of JPTA iter and mmFlexible.
This can be explained by studying the achieved beamforming gain sliced at the target angles $\theta^{(k)}|_{k=1,...,K}$ (eqn. (\ref{eqn:target_angles})), denoted by $\mathcal{B}_k(f_m)$ and defined as follows. 
\begin{equation}
\begin{aligned}
    \mathcal{B}_k(f_m)=G(\theta^{(k)},f_m)~~\forall~k=1,...,K, ~~\forall f_m
\end{aligned}    
\label{eqn:target_gain}
\end{equation}
where 
$G(\theta,f_m)$ is the beamforming gain function defined in (\ref{eqn:overall_gain}). Fig. \ref{fig:beam_gain_K2} and Fig. \ref{fig:beam_gain_K5} depict the on-target beamforming gain $\mathcal{B}_k(f_m)|_{k=1,...,K}$ for $K=2$ and $K=5$ users respectively, for $\{\theta_1,\theta_2\}=\{-30^\circ,40^\circ\}$.
For $K=2$, the average on-target gain achieved by Staircase TTD is lower than both mmFlexible and JPTA iter. However, when $K=5$, Staircase TTD achieves comparable on-target gain to both mmFlexible and JPTA iter. 
This is because the beam design methodology of Staircase TTD, which involves aligning the on-target-gain maxima with the respective sub-band centres, is not target-gain optimal for smaller $K(<4)$, and is hence outperformed by the optimization rooted mmFlexible and JPTA iter. 
However, a higher $K$ places stricter constraints on beam optimization, making the optimal solution converge to the beam design methodology of Staircase TTD as $K$ increases. This can be seen in  Fig. \ref{fig:beam_gain_K5}, where Staircase TTD not only achieves comparable average on-target gain to JPTA iter and mmFlexible, but also has its gain maxima aligned with those of JPTA iter and mmFlexible when $K=5$. 
This explains the observations made from Fig. \ref{fig:speff_vs_K}.  

Fig. \ref{fig:speff_vs_BW} and Fig. \ref{fig:speff_vs_N} study the effect of $BW$ and BS array size $N_T$, respectively, on the spectral efficiency for $K=5$ users. When $BW/f_c\leq 5\%$, Staircase TTD achieves comparable performance to both JPTA iter and mmFlexible. For $BW/f_c>5\%$, Staircase TTD is seen to exhibit greater robustness to beam squint effects compared to mmFlexible, but is outperformed by JPTA iter.  
In Fig \ref{fig:speff_vs_N}, Staircase TTD has comparable 
spectral-efficiency to both JPTA iter and mmFlexible for $N_T\leq 64$. Staircase TTD matches up to JPTA iter and considerably outperforms mmFlexible as $N_T$ increases thereafter. Fig. \ref{fig:speff_vs_SNR} shows that Staircase TTD achieves comparable performance to both JPTA iter and mmFlexible across SNRs for $K=5$, $N_T=32$, and $BW=2$GHz. Therefore, in summary, Staircase TTD achieves comparable performance to that of JPTA iter and mmFlexible when $K>4$, $BW/f_c\leq 5\%$ and $N_T\leq 64$, while outperforming mmFlexible when $BW/f_c>5\%$ and $N_T>64$. 

\section{Future work}
\label{sec:Future} 
While this work focuses on analog codebook design for sub-band-beam synthesis and theoretical performance evaluation in terms of spectral efficiency, our future work would study the practical challenges in RF front-end design to enable the prescribed sub-band-multiplexed multi-user data communication in realistic multi-user networks. In particular, we would study the impact of TTD hardware constraints, namely, delay range constraints \cite{delay_range}, limited phase shifter resolution, and non-linearity of circuit delays \cite{delay_error_1}, on the performance of sub-band-beams. In addition, a study of cross-sub-band interference and its mitigation is imperative for enabling sub-band-specific multi-user communication. Further, we would also study analog Staircase TTD codebooks with multi-stage frequency-spatial filtering, and multi-RF chain Staircase codebooks to realize beam patterns with arbitrary sub-band-angle mapping for highly flexible user-resource assignment. 

\section{Conclusions}
\label{sec:VI}
This paper proposes a structured, closed-form design of analog TTD codebook based on dual-stage frequency-spatial filter design to realize directional sub-band-beams to support simultaneous multi-user data communication. By implementing sub-band-selective filtering of directional grating lobes, it achieves beams with the required sub-band-angle mapping. It also delineates constraints on achievable sub-band-angle maps using the proposed codebook. The proposed method, besides espousing a conceptual visualization of sub-band-beam design, presents a low-cost and low-complexity analog TTD codebook design that matches the performance of optimization-rooted state-of-the-art approaches in large networks and exhibits reasonable robustness to beam-squint at large bandwidths. 
\appendices

{

}

\bibliographystyle{IEEEtran}
\bibliography{IEEEabrv,references}

\end{document}